\documentclass[a4paper]{article}
\usepackage{graphicx}
\usepackage{hyperref}
\usepackage{subcaption}
\usepackage{verbatim}
\graphicspath{{./images/}}
\usepackage{filecontents}
\usepackage[noadjust]{cite}
\usepackage{mathtools}

\usepackage{INTERSPEECH2019}

\title{End-to-End Bengali Speech Recognition}
\name{Sayan Mandal, Sarthak Yadav and Atul Rai}
\address{
  Staqu Technologies, India}
\email{sayan.mandal@staqu.com, sarthak.yadav@staqu.com, atul.rai@staqu.com}

\begin{document}
\maketitle
\begin{abstract}
  Bengali is a prominent language of the Indian subcontinent. However, while many state-of-the-art acoustic models exist for prominent languages spoken in the region, research and resources for Bengali are few and far between.
  In this work, we apply CTC based CNN-RNN networks, a prominent deep learning based end-to-end automatic speech recognition technique, to the Bengali ASR task. We also propose and evaluate the applicability and efficacy of small 7x3 and 3x3 convolution kernels which are prominently used in the computer vision domain primarily because of their FLOPs and parameter efficient nature. We propose two CNN blocks, 2-layer Block A and 4-layer Block B, with the first layer comprising of 7x3 kernel and the subsequent layers comprising solely of 3x3 kernels. Using the publicly available \textit{Large Bengali ASR Training data set}, we benchmark and evaluate the performance of seven deep neural network configurations of varying complexities and depth on the Bengali ASR task. Our best model, with Block B, has a WER of 13.67, having an absolute reduction of 1.39\% over comparable model with larger convolution kernels of size 41x11 and 21x11.

\end{abstract}

\section{Introduction}

Bengali, also referred to as \textit{Bangla}, is the seventh most spoken native language in the world, the second most widely spoken language in India and the official language of Bangladesh. It has around 185 million speakers in the Indian subcontinent. Bengali consists of 29 consonants, 7 vowels and 7 nazalized vowels and shares the phonetic space with other Magadhan languages like Assamese, Odiya etc. However, unlike various prominent languages spoken in the Indian subcontinent, research and resources for automatic speech recognition in Bengali are scarce.

With the emergence of voice as a natural form of human-computer interaction and the advent of large datasets, development of Automatic Speech Recognition (ASR) has seen major strides over the past few years in terms of robustness and practical applications. Deep Learning, facilitated by Convolutional Neural Networks (CNN) and Recurrent Neural Networks (RNN), as well as advancements in end-to-end training methodologies such as the Seq2Seq and Connectionist Temporal Classification paradigm have enabled the development of state-of-the-art end-to-end ASR pipelines.

 In this paper, we investigate the performance and applicability of existing deep learning based end-to-end automatic speech recognition paradigms, specifically CNN-RNN based hybrid models trained using Connectionist Temporal Classification (CTC) Loss in an end-to-end manner \cite{amodei2016deep} on the Bengali ASR task. We also perform ablation experiments to investigate the efficacy of smaller convolution kernels widely used in computer vision (\cite{simonyan2014very}, \cite{he2016deep}) which pose various benefits such as optimal number of parameters \cite{simonyan2014very} and lower memory addressing cost \cite{mashufflenet}, over the larger filters prevalent in the speech recognition domain \cite{amodei2016deep}. To this end, we benchmark and evaluate 7 different CNN-RNN based deep neural network architectures with various complexities and depth, on the \textit{Large Bengali ASR Training data set} \cite{kjartansson-etal-sltu2018}, a large scale publicly available Bengali corpus with around 200,000 utterances.

\section{Related Works}
\subsection{Deep learning and end-to-end ASR: Primer}
Traditional automatic speech recognition systems 
spearheaded by HMM-GMM based acoustic models, where HMMs were used for alignment and GMM mapped these aligned tri-phones to output characters, have dominated the field for quite some time, and only recently were surpassed by HMM-DNN systems, with deep neural networks replacing GMMs.

The advent of sequence-to-sequence models (Seq2Seq) paired with the development of Connectionist Temporal Classification (CTC) \cite{graves2006connectionist} revolutionized the end-to-end approach for training speech recognition systems by training a single network that directly mapped audio sequence to the output text, with recurrent neural network (\cite{graves2006connectionist}, \cite{hannun2014deep}, \cite{amodei2016deep}) and attention (\cite{chorowski2015attention}, \cite{chan2016listen}) based end-to-end models setting state-of-the-art results on several benchmark datasets.

Most recent deep learning based end-to-end systems utilize log-spectrogram or MFCC features as inputs to the network. Given the innate capability of deep neural networks to learn features from raw inputs, multiple recent works have switched to learning features from raw speech signal. To this end, \cite{Zeghidour2018} recently proposed time-delay convolution based trainable filterbanks, observing comparable performance with melfilterbank based systems on the WSJ dataset.

\subsection{Speech Recognition and Indian Languages}
Languages spoken in the Indian subcontinent are a topic of interest primarily due to the sheer number of speakers, but the lack of large-scale, publicly available datasets pose a challenge in the development and evaluation of deep learning based systems.

Most recently, a multilingual phone recognition system was proposed by \cite{KE2018}, covering four Indian langauges, viz. Kannada, Telugu, Bengali and Odia. They proposed a traditional HMM-DNN based phone recognition system leveraging Articulatory Features (AFs). Their DNN subsystem utilized tanh activation function at the hidden layers, and was trained in a greedy layer-by-layer supervised setting. However, unlike \cite{KE2018}, which proposed a phoneme recognition system evaluated on a very small dataset (19.2 hours total), we evaluate and propose CNN-RNN based end-to-end systems evaluated on a very large dataset.

Addressing the challenges involved in building robust speech recognition systems for Indian languages, \cite{Dash2018} proposed training phoneme recognition systems using articulatory movements captured by an electromagnetic articulograph. They evaluated multiple phoneme recognition models, similar to \cite{KE2018}, with the addition of HMM-RNN based acoustic model. However, they collected a small, 2 male speaker dataset for each language covered, which is an inconclusive sample size to comment on the genericity and applicability of the approach.

\section{Datasets}
\textit{Large Bengali ASR Training data set} (http://openslr.org/53/) \cite{kjartansson-etal-sltu2018} is used for evaluating models on the Bengali ASR task.
Since dev/test splits were not provided for the Bengali data set, we propose our own evaluation procedure, splitting the data set into train, validation and test sets (table.~\ref{tab:BengaliInfo}), available upon request.

\begin{table}[h]
  \caption{Bengali Dataset Statistics}
  \label{tab:BengaliInfo}
  \centering
  \begin{tabular}{c|c|c}
  	\toprule
	\textbf{Split} & \textbf{Utterances} & \textbf{Duration}\\
	\hline train & 148,110 & 145.90 hrs\\
	val & 26,138 & 25.79 hrs\\
	test & 43,562 & 42.91 hrs\\
  	\midrule
  	Total & 217,810 & 214.6 hrs\\
  	\bottomrule
  \end{tabular}
  
\end{table}

\section{Proposed Approach}
In the following subsections, we shed light on the evaluated CNN-RNN based end-to-end speech recognition networks. We also discuss the proposed changes to the original convolution as proposed in \cite{amodei2016deep}, as well as the various model architectures evaluated in this work.

The intention of evaluating multiple networks of varying depths and complexities are multi-fold:
\begin{enumerate}
\item To investigate performance trend when
	\begin{enumerate}
		\item Number of CNN layers are changed
		\item Number of RNN layers are changed
	\end{enumerate}
\item Benchmarking models for environments with different computational constraints
\end{enumerate}

\subsection{Model Outline}

\begin{figure}[t]
\begin{minipage}[b]{1.0\linewidth}
  \centering
  \centerline{\includegraphics[height=8cm]{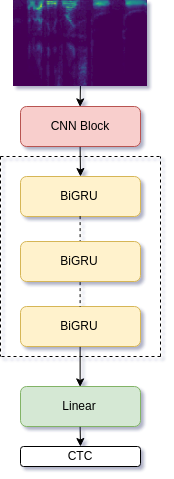}}
\end{minipage}	
\caption{Sample CNN-RNN Network for end-to-end ASR with 3 BiGRU layers}
\label{fig:cnnOutline}
\end{figure}

\begin{figure}[t]
\begin{minipage}[b]{1.0\linewidth}
  \centering
  \centerline{\includegraphics[height=4cm]{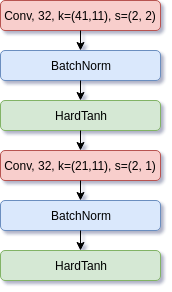}}
\end{minipage}	
\caption{Deep Speech 2 CNN Block}
\label{fig:originalBlock}
\end{figure}

Fig. \ref{fig:cnnOutline} depicts the general outline of the CNN-RNN based end-to-end speech recognition models that are popular in recent literature. (\cite{amodei2016deep}, \cite{hannun2014deep}). The convolution block represents \textit{n} sized stack of convolution layers. Fig. \ref{fig:originalBlock} shows the original 2-layer convolution stack as proposed in \cite{amodei2016deep}, where each convolution layer is followed by a Batch Normalization layer \cite{ioffe2015batch} and HardTanh non-linearity.

\subsection{Proposed Convolution Blocks}
The initial layers of a CNN based deep neural network are of significant importance in terms of overall network complexity as well as performance. The initial layers consume the most memory and comprise of the most floating point operations \cite{simonyan2014very} due to larger spatial dimensions of the input which is gradually downsampled with the depth of the network. 
Convolutional neural networks comprising solely of kernels with very small receptive fields (3x3 \cite{simonyan2014very}, 7x7 \cite{he2016deep}) have significantly pushed the state-of-the-art on various computer vision tasks and challenges such as ImageNet and COCO, and are widely used in recent works on cnn design. (\cite{he2016deep}, \cite{mashufflenet}, \cite{he2016identity}, \cite{zagoruyko2016wide}, \cite{szegedy2016rethinking}, \cite{huang2017densely}, \cite{szegedy2017inception}, \cite{huang2018condensenet}), and have various benefits over larger filters, viz,
\begin{enumerate}
\item Smaller kernels are more parameter efficient. \cite{simonyan2014very}
\item Multiple smaller kernels stacked over each other effectively increase the discriminative power of the decision function \cite{simonyan2014very}, by increasing the total number of non-linearities.
\item Larger convolution filters operated at a stride on the very first layer greatly reduce spatial resolution of filter maps, which might be too drastic and might hurt the performance of the network.
\end{enumerate}

\begin{figure}[t]
\begin{minipage}[b]{1.0\linewidth}
  \centering
  \centerline{\includegraphics[height=3cm]{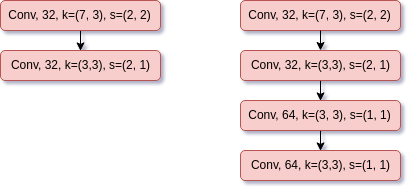}}
\end{minipage}	
\caption{Proposed CNN Blocks (batch normalization and hardtanh layers omitted)}
\label{fig:proposedBlock}
\end{figure}

We propose 2 convolution blocks with 2 and 4 convolution layers, henceforth referred to as Block A and B, respectively (Fig. \ref{fig:proposedBlock}). Instead of using large rectangular filters similar to \cite{amodei2016deep}, we utilize smaller 3x3 filters for all but the first layer (which uses 7x3 filters) in all our layers.

Table.\ref{tab:convFlops} depicts the total number of floating point operations (FLOPs) in millions and the number of parameters for each proposed convolution block as well as the 2-layer convolution block as proposed in DeepSpeech 2. As evident from the aforementioned table, the proposed convolution blocks have significantly lower FLOPs and parameters.

\begin{table}[t]
  \caption{Convolution Block Complexities}
  \label{tab:convFlops}
  \centering
  \begin{tabular}{c|c|c}
	\textbf{Block} & \textbf{FLOPs} & \textbf{Parameters}\\
	\hline DS 2\cite{amodei2016deep} & 1,640 M & 251.17 K \\
	\hline Block A & 69.80 M & 10.08 K\\
	\hline Block B & 398.1 M & 65.76 K\\
  \end{tabular}
\end{table}

\begin{table}[t]
  \caption{Evaluated Models}
  \label{tab:models_light}
  \centering
  \begin{tabular}{c|c|c}
  	\toprule
	Model & CNN Block & RNN Config\\
	\hline A-3GRU & Block A & 3, 512\\
	A-4GRU & Block A & 4, 512\\
	A-5GRU & Block A & 5, 512\\
	\midrule
	B-3GRU & Block B & 3, 512\\
	B-4GRU & Block B & 4, 512\\
	B-5GRU & Block B & 5, 512\\
	\midrule
	B-5GRU-Large & Block B & 5, 800\\
	\midrule
	2CNN-5GRU & DS 2 & 5, 800\\
	\bottomrule
  \end{tabular}
\end{table}

\subsection{Proposed Networks}
We propose and evaluate 7 network configurations of varying complexities as listed in Table \ref{tab:models_light}. 3 distinct RNN stacks are evaluated with 3, 4 and 5 bidirectional GRU layers respectively. Despite challenges in deployment in an online setting, bidirectional recurrent neural networks routinely outperform similar unidirectional models \cite{amodei2016deep}. Column \textit{RNN Config} depicts number of bidirectional layers, followed by the number of hidden units per layer. All configurations have 512 hidden units per bidirectional GRU layer, except \textit{B-5GRU-Large}, which has 800 hidden units.

\subsection{Implementation and Training Details}
Following \cite{amodei2016deep}, we use normalized log-spectrogram calculated using a sliding window of width 20ms and stride of 10ms, followed by a 160 point FFT as inputs to the network. The network is trained end-to-end using the CTC loss function \cite{graves2006connectionist}, which facilitates prediction of character sequences directly from input.

Training was performed on a single GTX 1080 Ti system with a batch size of 20 using Stochastic Gradient Descent, with an exponential learning rate decay schedule and an initial learning rate of $3e-4$. Early stopping \cite{caruana2001overfitting} with a patience of 3 was used for regularization, and all the networks were trained for a maximum of 30 epochs.

\textbf{Inference}: At inference time, we pair the proposed models with a 4-gram Kneser-Ney based language model with a beam size of 100, trained on a large corpus of Bengali text collected from Bengali news sites and blogs, using the Kenlm toolkit \cite{heafield2011kenlm}. Results without the language model are also provided for the test set.

\section{Experiments and Results}

Table~\ref{tab:BengaliResults} depict the word error rate (WER) for the Bengali ASR task on various networks on the val and test set. WER for validation is calculated without a language model, whereas results with and without language model are shown for the test set.

\begin{table}[h]
  \caption{Bengali ASR Results}
  \label{tab:BengaliResults}
  \centering
  \begin{tabular}{c c c c}
  \toprule
	\textbf{Model} & Val & Test (no LM) & Test\\
  \midrule
	A-3GRU & 37.7 & 37.86 & 14.99\\
	A-4GRU & 35.26 & 34.55 & 14.56\\
	A-5GRU & 37.76 & 36.75 & 14.24\\
	\midrule
	B-3GRU & 37.13 & 36.46 & 15.26\\
	B-4GRU & 34.10 & 33.51 & 14.50\\
	B-5GRU & 32.33 & 31.90 & 13.79\\
	\midrule
	B-5GRU-Large & \textbf{32.13} & \textbf{31.45} & \textbf{13.67}\\
	2CNN-5GRU & 34.71 & 33.74 & 15.06\\	
  \bottomrule
  \end{tabular}
\end{table}

Model B-5GRU-Large is the best performing configuration, achieving a WER of 13.67\%, followed closely by B-5GRU. It is worth noting that B-5GRU, despite having fewer hidden units per GRU layer as compared to 2CNN-5GRU (512 v/s 800) outperforms the latter by a significant margin: an almost 2\% absolute reduction in WER (without a language model), while at the same time consisting of a convolution block with lower number of parameters and FLOPs, hence validating the applicability of smaller convolution kernels for ASR task.

\section{Conclusions}
In this work, we investigated the performance and applicability of prominent end-to-end automatic speech recognition pipeline comprising of CNN-RNN based deep neural network trained using the CTC loss function for the Bengali language, evaluating and benchmarking various network configurations of different complexities. Our studies also validate the applicability of smaller convolution kernels widely used in the computer vision domain for ASR tasks, with the proposed convolution block A performing on par with convolution block comprising of larger 41x11 kernels keeping in consideration the drastically lower number of parameters for the former, whereas convolution block B outperforming the convolution block comprising of larger convolution kernels. The proposed network configurations can be applied to other Magadhan languages and comparable results can be expected as they share the phonetic space with Bengali.

\bibliographystyle{IEEEtran}
\bibliography{speakerbib}
\end{document}